\documentclass[11pt,a4paper,reqno]{amsart} 

\usepackage{amssymb,graphicx}

\usepackage{amssymb}

\newcommand{\koniec}{\begin{flushright}  $\Box $ \end{flushright}}
\newtheorem{theo}{Theorem}[section] 
\newtheorem{prop}[theo]{Proposition}

\def\theequation{\thesection.\arabic{equation}}

\newcounter{mnotecount}[section]

\renewcommand{\themnotecount}{\thesection.\arabic{mnotecount}}

\newcommand{\mnote}[1]
{\protect{\stepcounter{mnotecount}}$^{\mbox{\footnotesize
$
\bullet$\themnotecount}}$ \marginpar{
\raggedright\tiny\em
$\!\!\!\!\!\!\,\bullet$\themnotecount: #1} }

\newcommand{\hook}{{\setlength{\unitlength}{11pt}   
                   \begin{picture}(.833,.8)
                   \put(.15,.08){\line(1,0){.35}}
                   \put(.5,.08){\line(0,1){.5}}
                   \end{picture}}}
\newcommand{\CP}{\mathbb{CP}}
\newcommand{\C}{\mathbb{C}}

\newcommand{\R}{\mathbb{R}}
\newcommand{\HH}{\mathbb{H}}

\def\p{\partial}
\def\be{\begin{equation}}
\def\ee{\end{equation}}

\def\bea{\begin{eqnarray}}
\def\eea{\end{eqnarray}}

\def\ov{\overline}

\DeclareMathOperator{\tr}{Tr}

\usepackage{geometry}

\begin{document}
\date{April 18th, 2017}
\title{Manton's five vortex equations from self-duality}
\author{}
\author{Felipe Contatto }
\author{Maciej Dunajski}
\address{Department of Applied Mathematics and Theoretical Physics\\ 
University of Cambridge\\ Wilberforce Road, Cambridge CB3 0WA\\ UK.}
\email{felipe.contatto@damtp.cam.ac.uk, m.dunajski@damtp.cam.ac.uk}
\begin{abstract} 
We demonstrate that the five vortex equations recently introduced by Manton arise
as symmetry reductions of the anti-self-dual Yang--Mills equations in four dimensions. In particular
the Jackiw--Pi vortex and the Ambj\o rn--Olesen vortex correspond to the gauge group $SU(1, 1)$, and
respectively the Euclidean or the $SU(2)$ symmetry groups acting with two-dimensional orbits. We show how to obtain vortices with higher vortex numbers, by superposing vortex equations of different types.
Finally we use the kinetic energy of the Yang--Mills theory in 4+1 dimensions to construct a metric on vortex moduli spaces. This metric is not positive-definite in cases of non-compact gauge groups.
\end{abstract}   
\maketitle
\section{Introduction}
The Abelian Higgs model at critical coupling admits static solutions  modeling vortices 
which neither attract nor repel each other \cite{ManSutbook}. The mathematical  content of the model
consists of a Hermitian complex line bundle $L$ over a Riemannian surface $(\Sigma, g_\Sigma)$,
together with a $U(1)$ connection $a$ and a complex Higgs field $\phi$ satisfying the Bogomolny
equations
\[
\overline{\mathcal D}\phi=0, \quad f=\omega_\Sigma(1-|\phi|^2).
\]
Here $\omega_\Sigma$ is the K\"ahler form on $\Sigma$, 
$\overline{\mathcal D}=\overline{\p}-ia^{(0,1)}$ is the covariant $\overline{\p}$--operator (the anti-holomorphic part of the covariant derivative $D$ defined by $a$), $f=da$ is the Abelian Maxwell field and $a^{(0,1)}$ is the anti-holomorphic part of $a$.
Setting $|\phi|^2=\exp{(h)}$, and solving the first Bogomolny equation
for the one--form $a$ reduces the second equation to the Taubes equation
\cite{Taubes1980}
\be
\label{taubes}
\triangle h+2(1-e^h)=0,
\ee
which is valid outside small discs enclosing the logarithmic singularities of $h$ -- the locations
of vortices on $\Sigma$. Here $\triangle$ is the Laplace operator on $(\Sigma, g_\Sigma)$.

In \cite{Manton2017} Manton has considered a two--parameter generalisation of the Taubes equation
\be
\label{manton_eq}
\triangle h-2(C_0-Ce^h)=0.
\ee
The constants $C_0$ and $C$ can be rescaled to $0, 1$ or $-1$, and Manton has argued
that only five combinations lead to non--singular vortex solutions 
\begin{itemize}
\item $C=C_0=-1$ corresponds to the Taubes equation \cite{Taubes1980}. 
The magnetic field
$f$ decays to zero away from vortex  center.
\item $C=C_0=1$ is the Popov equation \cite{Popov2009, Popov2012}.
\item $C=0, C_0=-1$ corresponds to the magnetic field with constant strength
equal to $1$. In \cite{Manton2017} this was called the Bradlow vortex.
\item $C=1, C_0=-1$ is the Ambj\o rn--Olesen vortex. The magnetic
field is enhanced away from the position of the vortex \cite{AmbOl1988, Manton2017}.
\item $C=1, C_0=0$ is the Jackiw--Pi vortex equation, which arises
in a first order Chern--Simons theory \cite{JackPi1990, HorvZhang2009}. In this case $|\phi|^2$ tends
to zero at the position of the vortex and (non-compact surfaces) at $\infty$.
\end{itemize}
The aim of this paper is to show (Theorem \ref{main_theo} in \S\ref{sec_ec}) that Manton's equation 
(\ref{manton_eq}) for all values of $C_0, C$ arises as the symmetry reduction of the
anti-self-dual Yang--Mills equations (ASDYM) on a four-manifold
$M=\Sigma\times N$, where $N$ is a surface of constant curvature, and the symmetry group
is the group of local isometries of $N$. The value of $C_0$ in (\ref{manton_eq}) is determined
by the curvature of $N$, and $C$ depends on the choice of the gauge group 
$G_C$. We shall demonstrate that $N=S^2$ if $C_0=-1$, $N=\HH^2$ if $C_0=1$, and $N=\R^2$ if $C_0=0$.
The gauge group is $SU(2)$ if $C=-1$, $SU(1, 1)$ if $C=1$ and the Euclidean group $E(2)$ if $C=0$. In the integrable cases the Gaussian curvatures of 
$\Sigma$ and $N$ add up to zero. 

In \S\ref{sec_sup} we shall show how the five vortex equations are related by
a superposition principle which leads to a construction of vortices with higher vortex numbers.

The four-dimensional perspective allows us to derive a canonical expression for the resulting
energy of vortices. By considering the kinetic energy of the dynamical Yang--Mills theory on $\R\times M$ in \S\ref{sec_mec} we shall derive integral expressions for moduli space metrics corresponding to various choices
of constants in (\ref{manton_eq}). If the gauge group is non-compact, then the kinetic energy and the resulting moduli space metric are not positive definite and the moduli space may contain surfaces where they identically vanish. In the integrable cases of equation (\ref{manton_eq}), the moduli space metric takes a simple form that generalises the one for integrable Taubes vortices described in \cite{Strachan1992}. 

\subsubsection*{Acknowledgements} 
We thank Nick Manton and Edward Walton for helpful discussions. F.C. is grateful to Cambridge Commonwealth, European and International Trust and CAPES Foundation Grant Proc. BEX 13656/13-9 for financial support. M. D. was supported
by the STFC grant  ST/L000385/1.

\section{Equivariant anti-self-dual connections and symmetry reduction}
In this Section we shall formulate the main Theorem. Let us first introduce
some notation.
\subsection{The group $G_C$.} \label{secgroup}
The key role will be played by a Lie group $G_C\subset SL(2, \C)$ which consists of matrices $K$ such that
\[
K\left(\begin{array}{c c}
  1 & 0 \\ 
  0 & -C
 \end{array}\right)
K^\dagger=
\left(\begin{array}{c c}
  1 & 0 \\ 
  0 & -C
\end{array}\right), \quad\mbox{where}\quad C\in\R
\]
or equivalently
\be
\label{group_c}
G_C=\left\{K=\left(\begin{array}{c c}
  k_1 & k_2 \\ 
  C\overline{k_2} & \overline{k_1}
 \end{array}\right);k_1, k_2\in\mathbb C,\quad
 \text{and}\quad |k_1|^2-C|k_2|^2=1\right\}.
\ee
Therefore $G_{-1}=SU(2), G_{1}=SU(1, 1)$ and $G_0=E(2)$ -- the Euclidean group. 
The generators of the corresponding Lie algebra $\mathfrak{g}_C$,  
\be
\label{lie_alg_rep}
J_1=\frac{1}{2}\left(\begin{array}{c c}
  0 & i \\ 
  -Ci & 0
 \end{array}\right),\quad 	
 J_2=\frac{1}{2}\left(\begin{array}{c c}
  0 & -1 \\ 
  -C & 0
 \end{array}\right),\quad J_3=\frac{1}{2}\left(\begin{array}{c c}
  i & 0 \\ 
  0 & -i
 \end{array}\right),
\ee
satisfy commutation relations
\[
\left[J_1 ,J_2 \right]=-C J_3\; ,\quad \left[J_2, J_3\right]=J_1\; ,\quad \left[J_3 ,J_1 \right]=
J_2.
\]
The explicit parametrisation of $G_C$ as well as the left-invariant one-forms are constructed in the 
Appendix.	

In what follows, $G_{C_0}$ will denote the Lie group defined in the same way, but changing $C$ into $C_0$.

\subsection{The four-manifold} 
Let $M$ be the Riemannian four-manifold  given by the Cartesian product $\Sigma\times N$ with 
a product metric
\be
\label{metric_on_m}
g=g_\Sigma+g_N,
\ee
where $(\Sigma, g_\Sigma)$ is the Riemann surface introduced in \S{1}, 
and $(N, g_N)$ is a 
surface of constant Gaussian curvature $-C_0$. Let $w$ be a local holomorphic coordinate
on $\Sigma$, and $z$ be a local holomorphic coordinate on $N$
so that
\[
g_\Sigma=\Omega dw d\ov{w}, \quad \mbox{and}\quad
g_N=\frac{4 dzd\ov{z}}{(1-C_0|z|^2)^2},
\]
where $\Omega=\Omega(w, \ov{w})$ is the conformal factor on $\Sigma$.
The  K\"ahler forms $\omega_\Sigma$ on $\Sigma$  and  $\omega_N$ on $N$ are given by
\be\label{eqbeta}\omega_\Sigma= \frac{i}{2}\Omega dw\wedge d\ov{w}  , \quad
\omega_N=\frac{2i dz\wedge d\ov{z}}{(1-C_0|z|^2)^2}=2id\beta, \quad\mbox{where}\quad 
\beta=\frac{zd\ov{z}-\ov{z}dz}{2(1-C_0|z|^2)}.
\ee
We shall choose an orientation on $M$ by fixing the volume form $\mbox{vol}_M=\omega_\Sigma\wedge\omega_N$.
\subsection{Equivariance}
\label{sec_ec}
 Let $G_C$ and $G_{C_0}$ be Lie groups corresponding, via (\ref{group_c})
to two real constants  $C$ and $C_0$. In the Theorem below $G_C$ will play a role
of a gauge group, and $G_{C_0}$ will be the symmetry group.

Let $\mathcal E\rightarrow M$ be a vector bundle with a connection which, in a local trivialisation, 
is represented by a Lie-algebra valued one-form $A\in\mathfrak{g}_C\otimes\Lambda^{1}(M)$. The Lie
group $G_{C_0}$ is a subgroup of the conformal group on $(M, g)$, and acts on $M$ isometrically by
\begin{equation}\label{actionC}
(w, z)\mapsto \Big(w, \frac{k_1 z+k_2}{C_0\overline{k_2} z+\overline{k_1}}\Big).
\end{equation}
We shall impose the symmetry equivariance condition on $A$: it is preserved
up to a gauge transformation by the action (\ref{actionC}) of $G_{C_0}$.
The infinitesimal equivariance condition yields
\be
\label{equivariance_onA}
{\mathcal L}_X A= DW,
\ee
where ${\mathcal L}$ is the Lie derivative, $X$ is any vector field on $M$
generating the action (\ref{actionC}) and
$DW\equiv dW-[A, W]$ is the covariant derivative of a $\mathfrak{g}_C$-valued function on $M$.

In the coordinates $(w, \ov{w}, z, \ov{z})$ introduced above we have
\[
A=A_\Sigma+A_N, \quad \mbox{where}\quad A_{\Sigma}=A_wdw+A_{\ov{w}}d\ov{w}
\quad\mbox{and}\quad A_{N}=A_zdz+A_{\ov{z}}d\ov{z}.
\]
\newpage
\begin{theo}
\label{main_theo}
Let $A\in \Lambda^{1}(M)\otimes \mathfrak{g}_C$ be $G_{C_0}$--equivariant. Then
\begin{enumerate}
\item There exists a gauge  such that
\be
\label{Aansatz}
A=\left(\begin{array}{c c}
-C_0\beta+\frac{i}{2} a & -\frac{i}{1-C_0 z \overline z}\phi d\overline z\\ 
\frac{iC}{1-C_0 z \overline z}\overline\phi dz & C_0\beta-\frac{i}{2} a
\end{array}\right),
\ee
where $\beta$ is defined in (\ref{eqbeta}), $a$ is a $\mathfrak{u}(1)$-valued one-form, and $\phi$ is a complex Higgs field
on $\Sigma$.
\item The ASDYM equations on $(M, g)$ are
\be
\label{gen_bogomolny}
\overline{\mathcal D}\phi=0, \quad f+\omega_{\Sigma}(C_0-C|\phi|^2)=0
\ee
where $f=da$. Equivalently, setting $|\phi|^2=e^h$, 
\be
\label{manton_eq_thm}
\Delta_0 h-2\Omega (C_0-Ce^h)=0, \quad \mbox{where} \quad \Delta_0=4\p_w\p_{\ov{w}}.
\ee
\end{enumerate}
\end{theo}
We shall split the proof of this theorem into two Propositions
\begin{prop}
\label{prop1}
Let $G_{C_0}$ be the maximal group of isometries
of $(N, g_N)$, where $N=\R^2, S^2$ or $\HH^2$. 
The most general $G_{C_0}$--equivariant $G_C$--connection is gauge equivalent
to (\ref{Aansatz}).
\end{prop}
\noindent
and
\begin{prop} 
\label{prop2}
The ASDYM equations on (\ref{Aansatz}) are equivalent to (\ref{gen_bogomolny}) or 
(\ref{manton_eq_thm}).
\end{prop}
\noindent
{\bf Proof of Proposition \ref{prop1}.}
Every vector field $X$  generating the  $G_{C_0}$ action on $M$ corresponds to a $\mathfrak{g}_C$-valued function $\Phi_X$ on $M$ called the Higgs field. The symmetry group $G_{C_0}$ does not act freely on $N$, which leads to a set
of differential and algebraic constraints. These constraints are kinematical, as they arise purely from
the symmetry requirement, and do not involve the ASDYM equations. Our analysis of the
constraints, and the construction of $A$ follows 
\cite{ForgMan1980, ManSutbook}, and we refer the reader to these works
for details. Another method is presented in \cite{MasonWoodBook}.

Consider $N=G_{C_0}/U(1)$ as the homogeneous space, where $U(1)\subset G_{C_0}$ stabilises a point
with coordinates $z=0$ in $N$. This $U(1)$ action is generated by $J_3\in\mathfrak{g}_{C_0}$ defined in 
(\ref{lie_alg_rep}), and corresponds to a vector field $X^3$ on $M$. The constraint equations are
\be
\label{constraints_eq}
D\Phi_3=0, \quad [\Phi_3, \Phi_m]_{\mathfrak{g}_C}+
\Phi_{[X^3, X^m]_{\mathfrak{g}_{C_0}}}=0,
\ee
where the vector fields $X^m, m=1, 2, 3$ generate the $G_{C_0}$ action, and we set
$\Phi_m\equiv \Phi_{X^m}$. 
The map $X^3\mapsto\Phi_3$ is a homomorphism from $\mathfrak{u}(1)\subset\mathfrak{g}_{C_0}$
to $\mathfrak{g}_C$, and equations (\ref{constraints_eq}) imply that the $\Sigma$--components
of the potential $A_{\Sigma}=A_wdw+A_{\ov{w}}d\ov{w}$ belong to the image of this homomorphism
in the gauge Lie algebra, i.e.
$\mathfrak{u}(1)\subset\mathfrak{g}_{C}$. We also find that the components
$(A_w, A_{\ov{w}})$ are functions on $\Sigma$ which do not depend on the coordinates on $N$.
Therefore we have arrived at the Abelian Maxwell potential $a$ on $\Sigma$ given by
\[
a=a_wdw+a_{\ov{w}}d\ov{w}\equiv A_{\Sigma}.
\]
To solve (\ref{constraints_eq}) for the Higgs fields, we chose a gauge such that $\Phi_3$ is diagonal.
This is always possible if $C=-1$ as Hermitian matrices are diagonalisable by unitary changes
of basis, but needs to be verified by a direct calculation which uses the constraint equations
if $C=0$ or 
$C=1$. In general we find $\Phi_3=\epsilon J_3$, where $\epsilon=\pm 1$. The choice of $\epsilon$
amounts to a choice of an orientation\footnote{A calculation analogous to the one in this proof shows that if we choose $\epsilon=-1$, then the resulting form
 of $A$ will be
$$
A=\left(\begin{array}{c c}
  C_0\beta+\frac{i}{2}a & -\frac{i}{1-C_0 z \overline z}\phi dz\\ 
  \frac{iC}{1-C_0 z \overline z}\overline\phi d\overline z & -C_0\beta-\frac{i}{2}a
 \end{array}\right),
$$
which is equal to (\ref{Aansatz}) up to a change $z\leftrightarrow \overline z, w\leftrightarrow \overline w$.} 
on $\Sigma$, and we shall take $\epsilon=1$. The general
solution of (\ref{constraints_eq}) now becomes

\[
\Phi_1=\phi_1 J_1^{}+\phi_2 J_2, \quad
\Phi_2= \phi_2 J_1^{}-\phi_1 J_2, \quad
\Phi_3= J_3,
\]
where $\phi=\phi_2-i\phi_1$ is a complex valued function on $\Sigma$ which we shall later identify
with the Abelian Higgs field in the vortex equation.

 We shall now construct the remaining part of the gauge potential $A$ -- its components on
the surface $N$. Following \cite{ForgMan1980} define a gauge potential on $G_{C_0}$ by
\[
{\mathcal A}=\sum_{m=1}^3 \Phi_m \otimes \chi_m,
\]
where $\chi_1, \chi_2, \chi_3$ are right invariant one-forms on $G_{C_0}$ given by (\ref{l_forms}). In what follows, we use coordinates $(\kappa_1, \kappa_2, \kappa_3)$ for $G_{C_0}$ described in the Appendix.

We can perform a gauge transformation on $G_{C_0}$ such that
\[
\rho^3\hook \mathcal A=0, \quad {\mathcal L}_{\rho^3} \mathcal A=0,
\]
where $\rho^1, \rho^2, \rho^3$ are right-invariant vector fields on $G_{C_0}$ such that
$\rho^m\hook\chi_l={\delta^m}_l$, and $\rho^3=\p/\p \kappa_3$. 
This yields a gauge potential defined  on the quotient  $G_{C_0}/U(1)$, where the $U(1)$ subgroup is generated by $J_3$. It is given by
\[
\mathcal A=\mathcal A_1d\kappa_1+\mathcal A_2 d\kappa_2,
\]
where
\begin{align*}
\mathcal A_1&=-\phi_2 J_1+\phi_1 J_2\\
\mathcal A_2&=\frac{1}{\sqrt{-C_0}}\sin(\sqrt{-C_0} \kappa_1)\phi_1 J_1+\frac{1}{\sqrt{-C_0}}\sin(\sqrt{-C_0} \kappa_1)\phi_2 J_2+\cos(\sqrt{-C_0}\kappa_1)J_3.
\end{align*}
Notice that $\sin$ and $\cos$ above become hyperbolic functions for $C_0>0$. This one-form Lie derives along $\p/\p\kappa_3$ and thus we can use the diffeomorphism between $N$ and $G_{C_0}/U(1)$ to pull back $\mathcal A$ to $N$. This gives $A_N=\mathcal A$,
and finally $A=A_\Sigma+A_N$.  

A direct calculation shows that this gauge potential is indeed $G_{C_0}$-equivariant. Indeed, it satisfies (\ref{equivariance_onA}) where the vector fields $X^m, m=1, 2, 3,$ are push forwards of the  left-invariant vector fields
(\ref{vect_appendix})
 on $G_{C_0}$ by the projection $G_{C_0}\rightarrow N$,
and
\[
W_1=-\frac{\sqrt{-C_0}}{\sin(\sqrt{-C_0}\kappa_1)}\cos\kappa_2 J_3,\quad
W_2=-\frac{\sqrt{-C_0}}{\sin(\sqrt{-C_0}\kappa_1)}\sin\kappa_2 J_3,\quad
W_3	=0.
\]
In order to obtain a neater final expression, perform another gauge transformation by 
$\mbox{diag}(e^{-\frac{i}{2}\kappa_2},e^{\frac{i}{2}\kappa_2})$ and change into $z$-coordinates using the local
coordinate formula (\ref{zk_trans}) to find the $G_{C_0}$--equivariant $G_C$--gauge potential on $M$ given by (\ref{Aansatz}).
\koniec
\noindent
{\bf Proof of Proposition \ref{prop2}.}
Let $(w, z)$ be holomorphic coordinates on $M$. The basis of self-dual two forms
is spanned by \[
\Re(dw\wedge dz),\quad \Im(dw\wedge dz), \quad\omega_N+\omega_\Sigma.
\]
The ASDYM equations $(\omega_N+\omega_\Sigma)\wedge F=0$, and
$dw\wedge dz\wedge F=0$ take the form
\be
\label{asdym}
F_{wz}=0, \quad F_{\ov{wz}}=0, \quad \Omega^{-1} F_{w\ov{w}}+\frac{(1-C_0|z|^2)^2}{4}F_{z\ov{z}}=0.
\ee
The components of the gauge field $F_{\mu\nu}=\partial_\mu \mathcal A_\nu-\partial_\nu \mathcal A_\mu - [A_\mu, A_\nu]$ are given by
\begin{align*}
F_{z\overline z}&=\frac{-C_0+C\phi \overline\phi}{(1-C_0 z \overline z)^2}\,\sigma_3\,, \;\;\;&\;\;\; 
F_{w\overline w}&=\frac{i}{2}f_{w\overline w}\,\sigma_3 \,, \nonumber\\
F_{\overline z\overline w}&=\frac{i}{1-C_0 z \overline z}D_{\overline w}\phi\,\sigma_+\,,  \;\;\;&\;\;\
F_{z\overline w}&=-\frac{i}{1-C_0 z\overline z}D_{\overline w}\overline \phi\,\sigma_-\,, \nonumber\\
F_{zw}&=-\frac{i}{1-C_0 z \overline z}D_{y}\overline\phi\,\sigma_- \,,\;\;\;&\;\;\ 
F_{\overline z w}&=\frac{i}{1-C_0 z \overline z}D_{w}\phi\,\sigma_+\,,
\end{align*}
where $\phi=\phi_2-i\phi_1$, $f_{w \overline w}=\partial_w a_{\overline w}-\partial_{\overline w}a_w$, $D$ is the covariant derivative with respect to the $U(1)$-connection $a$ and
$$
\sigma_3=\left(\begin{array}{cc} 1 & 0 \\ 0  & -1 \end{array} \right)\,,\;\;\;\; \sigma_+=\left(\begin{array}{cc} 0 & 1 \\ 0  & 0 \end{array} \right)\,,\;\;\;\; \sigma_-=\left(\begin{array}{cc} 0 & 0 \\ C  & 0 \end{array} \right).
$$
Set
\[
D\phi =d\phi -i a\phi, \quad D\ov{\phi}=d\ov{\phi}+ia\ov{\phi}
\]
and
\[
\mathcal D=dw\otimes(\p_w-ia_w), \quad \ov{\mathcal D}=d\ov{w}\otimes(\p_{\ov{w}}-ia_{\ov{w}}), \quad
\mbox{so that} \quad D=\mathcal D+\ov{\mathcal D}.
\]
The ASDYM equations (\ref{asdym})
yield vortex-type equations
\begin{align}
D_{\overline w}\phi=\partial_{\overline w}\phi-ia_{\overline w}\phi=0,\label{vortex1}\\
\frac{i}{2} f_{w\overline w}+\frac{\Omega}{4}(-C_0+C \phi \overline\phi)=0.\label{vortex2}
\end{align}
This system of non--linear PDEs can be reduced to a single second order equation
for one scalar function. In fact, solve the first equation (\ref{vortex1}) for $a_{\overline w}$ so that $a_{\overline w}=-i\p_{\overline w}\ln(\phi)$ and, using the reality of $a$, $a_w=i\p_{w}\ln(\ov{\phi})$. Using these expressions for the components of $a$, calculate the Abelian Maxwell field $f_{w\overline w}$ and the second equation (\ref{vortex2}) yields (\ref{manton_eq_thm}).
\koniec
\subsection{Integrable cases} Following the integrability dogma \cite{MasonWoodBook,DunajskiBook, 
Cald2014}, a symmetry reduction of ASDYM is
integrable if the ASDYM equations are defined on a background $(M, g)$ with anti-self-dual Weyl curvature. Computing the Weyl tensor of (\ref{metric_on_m}) shows that conformal anti-self-duality is equivalent
to the vanishing of  the scalar curvature of $g$. Thus in the integrable cases the Riemann surface
$(\Sigma, g_\Sigma)$ on which the vortex equations are defined must have constant Gaussian curvature equal to minus the Gaussian curvature of $(N, g_N)$, i.e. locally,
\be\label{metricsM}
g_\Sigma=\frac{4 dwd\ov{w}}{(1+C_0|w|^2)^2}, \quad
g=\frac{4 dwd\ov{w}}{(1+C_0|w|^2)^2} + \frac{4 dzd\ov{z}}{(1-C_0|z|^2)^2}.
\ee
The local solutions of integrable vortex equations are given explicitly, in a suitable gauge, by \cite{Manton2017}
\be
\label{int_vortex}
\phi=\frac{1+C_0|w|^2}{1+C|s(w)|^2}\frac{ds}{dw},
\ee
where $s=s(w)$ is a holomorphic map from $\Sigma$ to a surface of curvature $C$. The vortices are located at zeros of $\phi$, which are the zeros of $ds/dw$ and the poles\footnote{Notice that $|\phi|^2$ is invariant under $s\mapsto 1/s$.} of $s$ of order at least $2$.

The integrable cases on simply-connected Riemann surfaces under the anti-self-duality framework are the following:
\begin{itemize}
\item The Taubes vortex ($C=C_0=-1$) is integrable on $\HH^2$, in which case it is a symmetry reduction from ASDYM on $\HH^2\times S^2$. In this case, $s$ is a Blaschke function \[
s(w)=\frac{(w-c_0)\dots(w-c_{\bf N})}{(w-\overline c_0)\dots(w-\overline c_{\bf N})},\] 
where $|c_k|<0$. This is the original integrable reduction of Witten \cite{Witten1977}.
\item The Popov vortex ($C=C_0=1$) is integrable on $S^2$, in which case it is a symmetry reduction from $S^2\times \HH^2$. In this case, $s:\CP^1\rightarrow\CP^1$ is a rational function $p(w)/q(w)$, where $p$ and $q$ are polynomials of the same degree with no common root.
\item The Bradlow vortex ($C=0$, $C_0=-1$) is integrable on $\HH^2$, in which case it is a symmetry reduction from $\HH^2\times S^2$.
\item The Ambj\o rn--Olesen vortex ($C=1$, $C_0=-1$) is integrable on $\HH^2$, in which case it is a symmetry reduction from $\Sigma\times S^2$.
\item The Jackiw--Pi vortex ($C=1$, $C_0=0$) is integrable on $\mathbb R^2$, in which case it is a symmetry reduction from $\mathbb R^2\times \mathbb R^2$. 
\end{itemize}
In each case the symmetry group is $G_{C_0}$ and the gauge group is $G_C$.

These integrable cases of (\ref{manton_eq}) do not exhaust the list of all
integrable vortices: there are other integrable cases
related to the sinh-Gordon and the Tzitzeica equations
\cite{Dunajski2012, ConDor2015, Contatto2017}, where the vortex
is interpreted as a surface with conical singularities.
\subsection{Superposition of vortices}
\label{sec_sup}
Given a solution to the vortex equation (\ref{manton_eq}) define the vortex number to be
\be
\label{vortex_number}
{\bf N}=\frac{1}{2\pi}\int_{\Sigma} f.
\ee
This is an integer equal to the first Chern number of the vortex line bundle $L\rightarrow \Sigma$,
and we shall assume that this integer is non--negative. 

Let us  now explain why there exist only five vortex equations among the nine possible combinations of values of $C$ and $C_0$. Equation (\ref{vortex1}) implies that 
the vortex number ${\bf N}$ coincides with the number of zeros of $\phi$ counted with multiplicities \cite{ManSutbook}. Since $\phi$ is holomorphic, ${\bf N}$ is necessarily non-negative. Since 
${\bf N}$ is proportional to the magnetic flux, the magnetic field $B\equiv-2if_{w\overline w}/\Omega$ can not be negative everywhere on $\Sigma$, but  (\ref{vortex2}) implies that this is only possible for the choice of constants 
\[(C,C_0)=(-1,-1), (0,-1), (1,-1), (1,0), (1,1), (0, 0),
\]
where the last possibility means that the magnetic field is null everywhere and (\ref{manton_eq}) is the Laplace equation. We shall call this the 
Laplace vortex.

The resulting six equations are not disconnected from one another. We shall  show that it is possible to construct higher order vortex solutions of one type by superposing two other types of vortices.  
Let $h$ be a vortex solution on $\Sigma$ with vortex number ${\bf N}$ satisfying 
\be
\label{bap_eq1}
\Delta_0 h +2\Omega (-C_0+C_1 e^h)=0,
\ee
so that $|\phi|^2=e^h$ has ${\bf N}$ isolated zeros, counting multiplicities. 
We say that this vortex is of type $(C_1, C_0)$.

Consider a metric
on $\Sigma$
\[
\tilde{g}_{\Sigma}=e^h g_{\Sigma}
\]
which has conical singularities at zeros of $|\phi|^2$.
Let  $\tilde h$ be a vortex solution with vortex number ${\bf \widetilde N}$ on 
$(\Sigma, \tilde{g}_\Sigma)$
satisfying 
\be
\label{bap_eq2}
\Delta_0 \tilde h +2 e^h \Omega (-C_1+C e^{\tilde h})=0,
\ee
so that  $|\phi|^2=e^{\tilde h}$ has ${\bf \widetilde N}$ zeros. We shall 
say that this vortex is of type $(C, C_1)$ with a rescaled metric. Adding both 
PDEs (\ref{bap_eq1}) and (\ref{bap_eq2}), we find
that 
\[
\Delta_0 (h+\tilde h) +2\Omega (-C_0+C e^{h+\tilde h})=0
\]
and that $e^{h+\tilde h}$ has ${\bf N}+\bf {\widetilde N}$ zeros. 
Therefore the superposition of a vortex of type $(C, C_1)$ with a rescaled metric on a vortex of type $(C_1, C_0)$ gives rise to a vortex of type $(C, C_0)$ with vortex number being the sum of the first two vortex numbers,
${\bf N}+{\bf \widetilde{N}}$. 
In the case $C=C_0=-1$ this is the Baptista superposition rule \cite{Baptista2014}.

Taking into account the six possible vortex equations, we make a list of all possible superpositions
\begin{align*}
\text{Taubes}+\text{Taubes}&=\text{Taubes}\\
\text{Bradlow}+\text{Taubes}&=\text{Bradlow}\\
\text{Laplace}+\text{Bradlow}&=\text{Bradlow}\\
\text{Ambj\o rn--Olesen}+\text{Taubes}&=\text{Ambj\o rn--Olesen}\\
\text{Jackiw--Pi}+\text{Bradlow}&=\text{Ambj\o rn--Olesen}\\
\text{Popov}+\text{Ambj\o rn--Olesen}&=\text{Ambj\o rn--Olesen}\\
\text{Popov}+\text{Jackiw-Pi}&=\text{Jackiw--Pi}\\
\text{Jackiw--Pi}+\text{Laplace}&=\text{Jackiw--Pi}\\
\text{Popov}+\text{Popov}&=\text{Popov},
\end{align*}
where the non-commutative summation $+$ means that the first vortex (of type $(C, C_1)$ with a rescaled metric) is superposed on the second one (of type $(C_1, C_0)$) to result in the vortex on the right hand side of the equality (of type $(C, C_0)$) with higher vortex number.
\section{Energy and Moduli Space Metric}
The energy functional $E$ of pure Yang-Mills theory on $M$ can be reduced to the energy function of an Abelian-Higgs model on $\Sigma$ using the ansatz (\ref{Aansatz}). This can be done by direct calculation,
\begin{align*}
E&=-\frac{1}{8\pi^2}\int_{M}\tr(F\wedge \star_g F)\\
&=\frac{1}{4\pi^2}\int_{N} \omega_N\int_\Sigma
\left[ \frac{1}{4}\left(B^2+(C_0-C\phi\overline\phi)^2\right)-\frac{C}{\Omega}
\left(\left|\mathcal D\phi\right|^2+\left|\overline{\mathcal D}\phi\right|^2\right)\right]\omega_\Sigma
\end{align*}
where \[
B=-2if_{w\ov{w}}/\Omega
\] 
is the magnetic field on $\Sigma$. This expression for the energy is proportional to the one given in \cite{Manton2017}.

If we assume that $N$ is compact\footnote{Notice that this is already the case if $C_0=-1$, as $N=S^2$. Otherwise we assume that the corresponding surfaces are quotiented out by a discrete subgroup of $G_{C_0}$. If $C_0=0$, $N$ is a two-torus $\mathbb T^2$ ($\mathbb R^2$ quotiented out by a lattice) and if $C_0=1$ $N$ is a compact Riemann surface of genus ${\tt g}>1$ ($\mathbb H^2$ quotiented out by a Fuchsian group). The ansatz (\ref{Aansatz}) must then admit
this further discrete symmetry.} then the  first integral is the area of $N$ given by
$$
\mbox{Area}_N=\int_{N} \omega_N=\begin{cases}
      \frac{4\pi}{-C_0}, & \text{if}\ C_0<0 \\
      4\pi, & \text{if}\ C_0=0 \\
      \frac{4\pi}{C_0}({\tt g}-1) & \text{if}\ C_0>0,
    \end{cases}
$$ 
where ${\tt g}$ is the genus of $N$ and we normalised the area of the torus ($C_0=0$) to $4\pi$.
Thus the energy can be written, using the Bogomolny argument along with $[D_w,D_{\overline w}]\phi=-if_{w\overline w}\phi	$ and the integration by parts
(with an additional boundary condition $D\phi=0$ if $\Sigma$ is not compact) as
$$
E=\frac{\mbox{Area}_N}{16\pi^2}\int_\Sigma\left[\left(B+C_0-C|\phi|^2\right)^2-\frac{8C}{\Omega}\left|\overline{\mathcal D}\phi\right|^2\right]\omega_\Sigma-C_0\frac{\mbox{Area}_N}{4\pi}{\bf N},
$$
where ${\bf N}$
is the vortex number defined by (\ref{vortex_number}).
If the vortex equations (\ref{vortex1}) and (\ref{vortex2}) are satisfied, then the energy is proportional to the vortex number, characterising a non-interacting theory, and this value is the global minimum of $E$ if $C\leq 0$. However, equation (\ref{vortex1}) cannot be naturally derived from this argument if $C=0$. In fact the theory corresponding to this energy functional does not involve any Higgs field in this case even though the symmetry reduction of ASDYM necessarily gives rise to a holomorphic Higgs field satisfying (\ref{vortex1}). This is a counter-example to the principle of symmetric criticality, proved under certain conditions in \cite{Palais1979}.

If $N$ is not compact we can still make sense of {\it energy density} (or energy per unit of area of $N$).

Originally, $E$ is the energy functional of pure Yang-Mills theory in four dimensions, which under the ASDYM condition is 
$$
E=\frac{1}{8\pi^2}\int_{\mathcal M}\tr\left(\mathcal F\wedge\mathcal F\right)\equiv k,
$$
where $k$ is the instanton number. 
Comparing both expressions for the energy, we derive a relation between the vortex and instanton numbers,
$$
k=-C_0\frac{\mbox{Area}_N}{4\pi}{\bf N}=(1-{\tt{g}}){\bf N}, \quad {\tt g}=0,1,2,\dots\,.
$$
\subsection{Dynamical theory}
\label{sec_mec}
Yang--Mills instantons on $M$ can be regarded as static solitons on $\R\times M$ with a product
metric $-dt^2+g$. Implementing the symmetry reduction of Theorem \ref{main_theo}, but from five dimensions, leads to vortices on $\Sigma$ interpreted as stationary solitons in a dynamical theory on 
$\R\times\Sigma$. We shall use this approach to find the kinetic term in the total energy 
functional on $\R\times\Sigma$, and use it to read-off the metric on the moduli space
of static vortices. Let $\mathcal F$ be a $\mathfrak{g}_C$--valued Yang--Mills field
on $\R\times M$.
The action functional of pure YM theory with $t$-dependence is
$$
S=-\frac{1}{8\pi^2}\int_{\mathbb R\times M}\tr(\mathcal F\wedge\star_5\mathcal F)=\int_{\mathbb R}\mathcal L dt,
$$
where $\mathcal L$ is defined by the second equality and involves the integral on $M$ alone. Under the symmetry reduction of Theorem \ref{main_theo}, $\mathcal L$ becomes a Lagrangian on $\R\times\Sigma$,
$$
\mathcal L= -\frac{\mbox{Area}_N}{4\pi^2}
\int_\Sigma
\Big(\frac{1}{4}(B^2+(C_0-C\phi\overline\phi)^2)-\frac{C}{\Omega}
(D_{\overline w}\overline\phi D_w\phi+D_{\overline w}\phi D_w\overline\phi) -\frac{1}{\Omega} f_{0w}f_{0\ov{w}}+\frac{C}{2}|D_0\phi|^2\Big)\omega_\Sigma.
$$
The Euler-Lagrange equations, resulting from calculating the variation with respect to $\phi, a_w$ 
and $a_0$, are, respectively,
\begin{align}
D_wD_{\overline w}\overline\phi+D_{\overline w}D_{w}\overline\phi-\frac{\Omega}{2}D_0D_0\overline\phi+\frac{\Omega}{2}(-C_0+C \phi\overline\phi)\overline\phi=0,\label{EL1}\\
-2\partial_{\overline w}\left(\frac{1}{\Omega}f_{w\overline w}\right)+\partial_0f_{0\overline w}-iC(D_{\overline w}\overline\phi \phi-D_{\overline w}\phi \overline\phi)=0,\label{EL2}\\
\partial_wf_{0\overline w}+\partial_{\overline w}f_{0w}+\frac{iC\Omega}{2}\left(\phi D_0\overline\phi-\overline\phi D_0\phi\right)=0.\label{EL3}
\end{align}
The equations resulting from varying $\overline\phi$ and $a_{\overline w}$ are the complex conjugate of equations (\ref{EL1}) and (\ref{EL2}), respectively. The third equation is usually referred to as Gauss' law.

This system of  second order dynamical equations is satisfied by static solutions to the first order vortex equations (\ref{vortex1}--\ref{vortex2}) in the temporal gauge $a_0=0$. In fact, Gauss' law (\ref{EL3}) is automatically satisfied. To see that (\ref{EL1}) is satisfied, use (the complex conjugate of) (\ref{vortex1}) to write $D_wD_{\overline w}\overline\phi=[D_w, D_{\overline w}]\overline\phi=if_{w\overline w}\overline\phi$ and eliminate $f_{w\overline w}$ with (\ref{vortex2}). Finally, equation (\ref{EL2}) is satisfied upon eliminating $f_{w\overline w}$ in the same way and using $\partial_{\overline w}\left(\phi\overline\phi\right)=
(D_{\overline w}\phi)\overline\phi+\phi D_{\overline w}\overline\phi=\phi D_{\overline w}\overline\phi$.

The kinetic energy $T$ can be read off from $\mathcal L$. In the  temporal gauge
when $a_0=0$ it takes the form
\be
\label{kin_en}
T=\frac{\mbox{Area}_N}{8\pi^2}\int_\Sigma\left[\frac{2}{\Omega}\dot a_w \dot a_{\overline w}-C\dot\phi \dot{\overline\phi}\right]\omega_\Sigma,
\ee
where the dots denote $t$-derivatives. This generalises the known kinetic energy for the Taubes vortex ($C,C_0<0$) \cite{Strachan1992}.

In the usual Abelian Higgs model in the critical coupling (yielding Taubes vortices), there is a moduli space ${\mathcal M}_N$ 
of static ${\bf N}$ vortex solutions. All these 
solutions have the same potential energy, so there are no static forces. The moduli space
acquires a metric from the kinetic energy of the theory, and the geodesics of this metric 
model slow motion of ${\bf N}$-vortices. There are several ways to obtain the metric for both 
flat and curved backgrounds \cite{ManSutbook,DorDunMan2013}. One way to proceed 
for the integrable vortices  is to assume that the vortex positions 
depend on time, and
substitute the explicit solution (\ref{int_vortex}) into the kinetic energy (\ref{kin_en}).
This, when quotiented out
by the gauge equivalence (which is equivalent to imposing Gauss' law \cite{ManSutbook}), gives a quadratic form on ${\mathcal M}_N$. In case of Taubes vortices
the resulting metric is positive definite, but we see that (\ref{kin_en}) is not positive definite
if $C>0$, which is the case for Jackiw--Pi, Popov and Ambj\o rn--Olesen vortices. 

Samols derived a localised expression for the metric of the moduli space of Taubes vortices \cite{Samols1992} (see also \cite{Strachan1992} 
and \cite{martin}
for the metric of moduli space of hyperbolic vortices), where the moduli are the vortex positions (or zeros of the Higgs field). The moduli space metric of Taubes vortices with simple zeros $\{W_1,\dots,W_{\bf N}\}$ is, from (\ref{kin_en}),
$$
\frac{Area_N}{8\pi}\sum_{i,j=1}^{\bf N}\left(\Omega(W_k)\delta_{ij}+2\partial_{W_i}b_j\right)dW_i d \overline W_j,
$$
where $b_j=\partial_{\overline w}(h-2\log|w-W_j|)|_{w=W_j,\overline w=\overline W_j}$ and $\Omega(W_i)$ means that the conformal factor $\Omega$ is being evaluated at the point $(w,\overline w)=(W_i,\ov W_i)$.

A similar calculation as the one performed by Samols (cf. also \cite{ManSutbook}) adapted to vortices defined by (\ref{manton_eq}) gives the following result for the metric on the moduli subspace associated to the simple zeros of the Higgs field
\be\label{metrticmoduli}
\frac{Area_N}{8\pi}\sum_{i,j=1}^{\bf N}\left(-C_0\Omega(W_k)\delta_{ij}+2\partial_{W_i}b_j\right)dW_i d\overline W_j,
\ee
where the constant $C_0$ appears because the calculation involves the use of equation (\ref{manton_eq}).

In the integrable case, the background metric is locally given by $g_\Sigma$ in (\ref{metricsM}) and the Higgs field is (\ref{int_vortex}) in a particular gauge. As above, we assume that $|\phi|$ admits only simple zeros and that each of them is as zero of $ds/dw$ and not a pole, which is always the case up to performing the transformation $s\mapsto 1/s$. In this case, $b_j$ can be calculated directly and is given by
$$
b_j=C_0 \frac{W_j}{1+C_0\left|W_j\right|^2}+\frac{3}{2}\overline\beta_j,
$$
where $\beta_j=\frac{s_3^{(j)}}{s_2^{(j)}}$ and $s^{(j)}_k=k!\dfrac{d^ks}{dw^k}\big|_{w=W_j}$, $k=0,1,2,3,\dots$, that is to say, the  $s^{(j)}_k$'s are defined by
$$
s(w)=s_0^{(j)}+s_2^{(j)}(w-W_j)^2+s_3^{(j)}(w-W_j)^3+\cdots,
$$
where the linear term is absent because we assumed that $W_j$ is a simple zero of $s$.

Now, the moduli space metric (\ref{metrticmoduli}) of simple vortices can be calculated and is given by
\be\label{metricmoduliint}
\frac{3\, Area_N}{16\pi}\sum_{i,j=1}^{\bf N}\dfrac{\partial \overline\beta_j}{\partial W_i} dW_i d \overline W_j.
\ee
This is in agreement with the formula derived for integrable Taubes vortices in \cite{Strachan1992}. In particular, (\ref{metricmoduliint}) tells us that the metric is zero when $s$ depends only holomorphically on the vortex positions, which is the case for some Popov and Jackiw--Pi vortices. In fact, this is what happens for the ${\bf N}=2$ Popov vortex on $\Sigma=S^2$ corresponding to $s=(w-W_1)^2/(w-W_2)^2$, and to the ${\bf N}=1$ Jackiw--Pi vortex on $\Sigma=\mathbb R^2$ corresponding to $s=(w-W_1)^2$.

\section*{Appendix. The group $G_{C_0}$}
\appendix
\setcounter{equation}{0}
\label{Appendix_sec}
\def\theequation{\thesection{A}\arabic{equation}}
In this Appendix, we denote by $J_m^{C_0}$, $m=1,2,3$ the generators of the Lie-algebra $\mathfrak g_{C_0}$ given by (\ref{lie_alg_rep}) with $C$ replaced by $C_0$.

 A parametrisation of $G_{C_0}$ is given by 
$$
K=\left(\begin{array}{c c}
  e^{i(\kappa_3-\kappa_2)/2}\cos(\sqrt{-C_0}\kappa_1/2) & -\frac{1}{\sqrt{-C_0}}e^{i(\kappa_2+\kappa_2)/2}\sin(\sqrt{-C_0}\kappa_1/2) \\ 
  \sqrt{-C_0}e^{-i(\kappa_3+\kappa_2)/2}\sin(\sqrt{-C_0}\kappa_1/2) & e^{-i(\kappa_3-\kappa_2)/2}\cos(\sqrt{-C_0}\kappa_1/2)
 \end{array}\right),
$$
where $0\leq\kappa_3\leq 4\pi$, $0\leq\kappa_2\leq 4\pi$, $0\leq\kappa_1<\pi/\sqrt{-C_0}$ if $C_0<0$ and $\kappa_1\geq 0$ if $C_0\geq 0$.

The coordinate $\kappa_3$ parametrises the $U(1)$ fibres of the fibration
$G_{C_0}\rightarrow N=G_{C_0}/U(1)$. In the proof of Proposition \ref{prop1} we  need expressions relating the
local coordinates $(z, \ov{z})$ on $N$ to $(\kappa_1, \kappa_2)$ on 
$G_{C_0}/U(1)$. Let $p\in N$ be a point corresponding
to the coordinate $z=0$. Consider the group action (\ref{actionC}) such that the RHS is $0$. This
gives a system of two equations for $(z, \ov{z})$ with a solution
\be
\label{zk_trans}
z=\frac{1}{\sqrt{-C_0}}\tan(\sqrt{-C_0} \kappa_1/2)e^{i\kappa_2}.
\ee
Note that $\frac{1}{\sqrt{-C_0}}\tan(\sqrt{-C_0} \kappa_1/2)\geq 0$ regardless of the sign of $C_0$. 
The formula (\ref{zk_trans}) is  well defined for $C_0=0$ upon taking the limit $C_0\to 0$. The coordinate $\kappa_3$ of $G_{C_0}$ parametrises the stabiliser of $p\in N$, which is a $U(1)$ subgroup generated by $J_3^{C_0}$.

The right-invariant one-forms $\chi_1,\chi_2,\chi_3$  such that
$(dK) K^{-1}+\sum_{m=1}^3 \chi_m \otimes  {J_m^{C_0}}=0$,
and the left--invariant vector fields
$\eta^1, \eta^2, \eta^3$ on $G_{C_0}$ are given by
\begin{align}
\label{l_forms}
\chi_1=&\left(\frac{1}{\sqrt{-C_0}}\sin(\sqrt{-C_0} \kappa_1)\cos\gamma d\kappa_2+\sin\kappa_3 
d\kappa_1\right)\nonumber\\
\chi_2=&\left(-\cos\kappa_3 d\alpha+\frac{1}{\sqrt{-C_0}}\sin\gamma \sin(\sqrt{-C_0}\,\kappa_1)
d\kappa_2\right)\nonumber\\
\chi_3=&\left(-d\kappa_3+\cos(\sqrt{-C_0}\,\kappa_1)d\kappa_2\right),
\end{align}
and
\begin{align}
\label{vect_appendix}
\eta_1&=-\sin{\kappa_2}\partial_{\kappa_1}-\frac{\sqrt{-C_0}}{\tan(\sqrt{-C_0}\kappa_1)}\cos\kappa_2 \partial_{\kappa_2}-\frac{\sqrt{-C_0}}{\sin(\sqrt{-C_0}\kappa_1)}\cos\kappa_2\partial_{\kappa_3},
\nonumber\\
\eta_2&=\cos\beta\partial_{\kappa_1}-\frac{\sqrt{-C_0}}{\tan(\sqrt{-C_0}\kappa_1)}\sin\kappa_2 
\partial_{\kappa_2}-\frac{\sqrt{-C_0}}{\sin(\sqrt{-C_0}\kappa_1)}\sin\kappa_2\partial_{\kappa_3} 
\nonumber\\
\eta_3&=-\partial_{\kappa_2}.
\end{align}

\bibliography{BiblioArticle}
\bibliographystyle{plain}

\end{document}